# Pan-tropical plant functional trait variation from space

To be submitted to New Phytologist

David Schimel[1], Andres Baresch[2,3], Adam Chlus[1], Phil Townsend[3], Fabian Schneider[1,4], Gaia Vaglio Laurin[5] and Ben Poulter[2,7]

1) Jet Propulsion Lab, California Institute of Technology, Pasadena, CA 91101
2) University of Maryland, Department of Geographical Sciences, College Park, MD 20742
3) Goddard Space Flight Center, 8800 Greenbelt Road, Greenbelt, Maryland 20771
4) University of Wisconsin, Russell Labs, 1630 Linden Drive, Madison, WI 53706
5) Aarhus University, Ny Munkegade 116, 8000 Aarhus C Denmark
6) Research Institute on Terrestrial Ecosystems, National Research Council, Montelibretti, Italy and PRISMA Science Team,
7) Spark Climate Solutions, San Francisco, CA

## Summary

Introduction: 960 words, materials and methods: 1880 words, results and discussion: 1682 words, 2021 words, conclusions 709, 2 Color Figures in main text, 8 figures total, 2 tables, Supporting information, figures (3 figures).

## Summary

- Plant functional trait variation in tropical moist forests is central to predicting ecosystem responses to change. Information on traits in the tropics is limited relative to the diversity of climates, landforms, disturbance regimes and species present. These traits are key to modeled predictions of ecosystem change.
- We used a spaceborne imaging spectrometer from the Italian Space Agency to sample roughly 1% of the tropical moist forest biome for traits along the Leaf Economic Spectrum, as well as data to contrast them to adjacent tropical biomes. We used this data set to examine LES traits variation between tropical moist forests on three continents.
- Tropical forests in Asia, Africa and the Americas and across Wallace's line occupy different parts of trait space. Topographic complexity emerged as a control, possibly reflecting both local adaptation and orographic isolation. Tropical forests should not be treated as functional replicates on three continents.
- This study creates a baseline for future advanced spectroscopic sensors to monitor a wider range of plant functional traits and changes over seasonal and longer time scales, Knowledge of trait variation, and its environmental can inform models of ecosystem response in a changing environment, allowing biological detail in models of biophysical and biogeochemical processes.

## Introduction

The responses of ecosystems in a changing environment depends on the spectrum of variation in organismal responses, defining the range of responses accessible in a region or biome. This requires understand the range of plant functional traits, which can vary widely and are challenging to sample adequately. Traits related to growth constrain energy, water and carbon exchange and reflect nutrient status and so are crucial for modeling (Schimel *et al.*, 2019; Braghiere *et al.*, 2024; Laurin *et al.*, 2024; Baresch *et al.*). Sampling is particularly challenging for tropical forest ecosystems, which contain a vast number of species with comparatively few observations to characterize variation at species, community, landscape and continental scales (Schimel *et al.*, 2015; Asner *et al.*, 2017a).

Tropical forests have long been classified a single biome and modeled as one or a small number of functional types (Levine *et al.*, 2016). Models and many other analyses simplify the range of possible responses partly because parameterizing these responses is limited to data that data are sparse relative to the immense variation found there. Data to characterize traits and trait-environment relationships in the tropics are also limited relative to other regions and even more so considering the degree of variation relative to mid-and-high latitude forests (Schimel *et al.*, 2015). These sparse and non-representative data on tropical forest traits limits understanding how the three floras function and differ in function (Schimel *et al.*, 2015).

Despite the tendency to model the tropical regions on continents as replicates, the value of cross-continental comparisons have been emphasized in the literature and recent studies reveal differences in ecosystem function (Liu *et al.*, 2017; Bennett *et al.*, 2021). While significant efforts have been made to identify mapped covariates to use in upscaling, data to calibrate relationships is very sparse and unevenly distributed. Local studies show complex trait-environment relationships, suggesting upscaling from limited data over complex landscapes could be misleading (Messier *et al.*, 2010; Chadwick & Asner, 2020).

Regional environmental gradients, local soils and topographic variation between sites and species differences within sites all play a significant role on traits (Kreft & Jetz, 2007; Kraft & Ackerly, 2010; Messier *et al.*, 2010; Chadwick & Asner, 2020). Fine-scale gradients nested within larger climatic gradients can shift the traits expected by considering climate zones based on climate alone and available covariates may not capture smaller scales in any case (Barton & Fortunel, 2023). As an example, the influence of soil variation, hillslope hydrology, herbivory, disturbance and other factors could vary between mountainous Asia and the more subdued topography of the Amazon basin. Heterogeneity can play a major role in ecosystem response to change (Levine *et al.*, 2016). Remotely observed traits (Asner *et al.*, 2017a; Cavender-Bares *et al.*, 2022) are emerging to play a complimentary role to field studies sampling variation over landscape, orographic/topographic and larger gradients (Kreft & Jetz, 2007).

Several lines of evidence suggest that ecosystem function may very between continental tropical regions found in the Asia-Pacific, Africa and the Americas. These regions differ in many climatic and topo-edaphic features but are often treated as replicates in models (Ordway *et al.*, 2022). Liu et al (Liu *et al.*, 2017) showed that these regions varied in their response to the 2015-2016 El Nino event, and it has been argued that Africa is more drought-resilient because paleo-drought events have selected there for drought-resistant species. In their comprehensive review, Primack and Corlett describe a plethora of biological and evolutionary differences (Corlett & Primack, 2011).

Remote observations allowing sampling within a defined time period, using a consistent method, sampling much larger areas and avoiding bias due to site selection or accessibility, and can provide information on aspects of biodiversity (Vaglio Laurin *et al.*, 2014; Schneider *et al.*, 2020; Braghiere *et al.*, 2024). We take advantage of the PRISMA satellite launched by the Italian Space Agency in 2019 carrying onboard an imagining spectrometer, to collect hyperspectral imagery over tropical areas (Fig. 1) and quantifying the (LES)(Wright *et al.*, 2004), avoiding the sampling limitations in situ (Schimel *et al.*, 2015).

In Africa and Asia, (Aguirre-Gutiérrez *et al.*, 2025) used data from 657 ha in the Americas, 126 in Africa and 16 ha in Asia or about 8 sq km in which foliage was sampled. By contrast in in the Americas, PRISMA sampled ~80, 000 sq km of canopies, in Asia, ~50,000 sq km and in Africa ~35,000 sq km or 165,000 sq km. Supplemental Fig. 3 shows the distribution of these scenes in geographical and climate space. Rocchini et al (2022) describes potential new approaches to biodiversity science based on massive sampling with remote

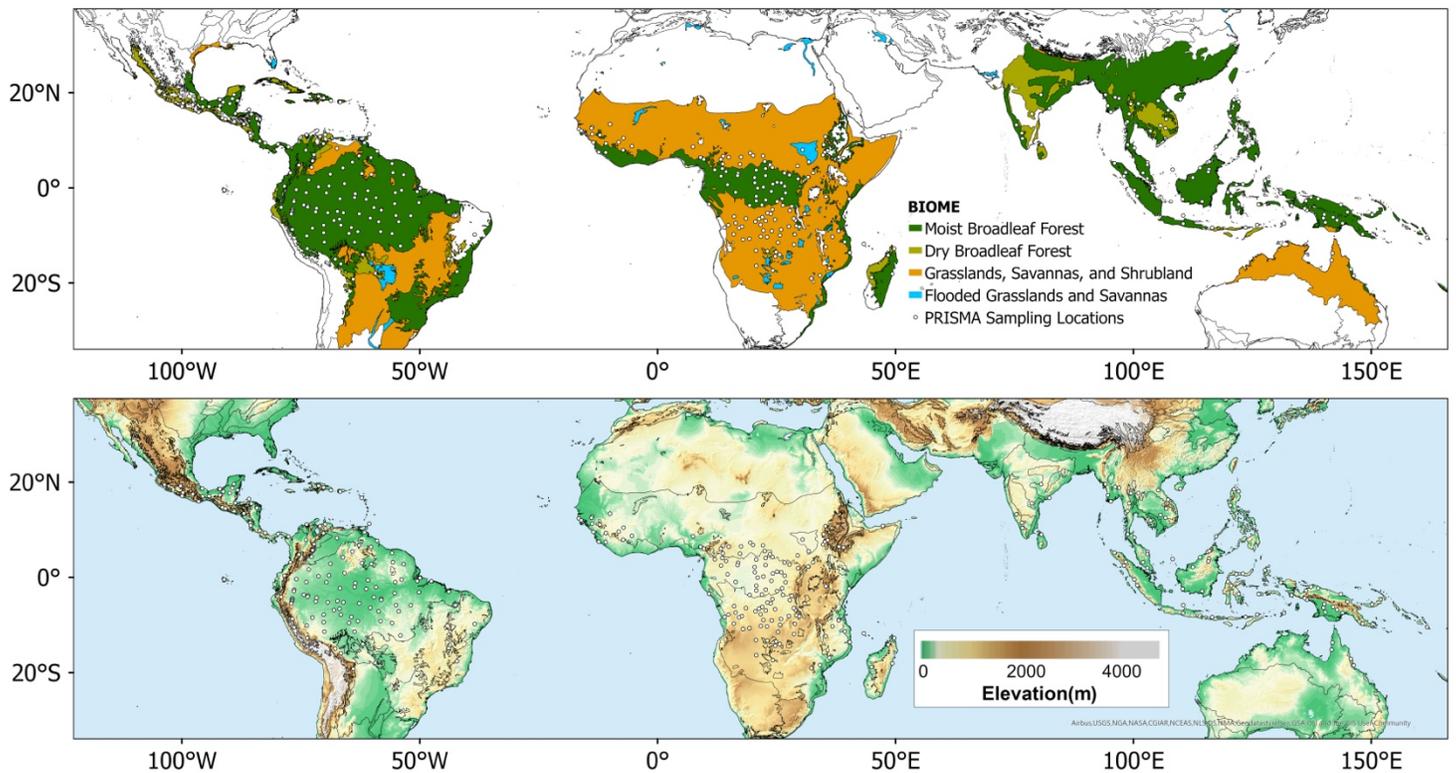

sensing: here we provide an early demonstration.

*Figure 1. PRISMA samples the tropics with a random protocol (although some sites were requested by specific investigators). Tropical biomes* (Buchhorn *et al.*, 2020) *based on Olson (1983) for tropical regions on each continent and the locations of PRISMA 30 x 30 km scenes shown over topography. The Americas have 88 scenes in tropical most forest, Africa*

*has 38 and Asia 58. Each scene covers 900 sq km and in total, the data cover about 1% of tropical moist forest. Adjacent tropical biomes are shown for context. See Supplemental Figure S3 for the PRISMA scenes in geographic and climate space.*

Remote sensing is a demonstrated method for traits retrievals in the tropics, and the LES captures traits combinations of plants along the gradient from fast to slow growing strategies, and is often indexed by the evolutionary constraints between foliar nitrogen, chlorophyll content and leaf dry mass per unit area. High nitrogen and chlorophyll leaves allow high rates of photosynthesis and rapid growth, while thicker (and heavier) leaves require more invested resources and are typical of slower growing species (Wright *et al.*, 2004). The LES can be retrieved for the pan-tropics using an instrument with PRISMA's characteristics.

If traits distributions vary strongly along steep environmental gradients, soils, microclimate, elevation or disturbance, or between regions separated by mountains, then a small sample, even if random, may not capture variation within the ecosystem and local variation could alias estimates of larger scale patterns. This is an issue of concern in parts of the tropics with strong orographic or hydrological patterning. Sampling over environmental gradients within tropical regions is critical to get an unbiased picture of continental differences. The pool of functional diversity available to respond to environmental change will draw on the full range of variation, and not just those aspects of variation represented along macro-environmental gradients.

Remote sensing can provide data and vastly more data than extant trait data bases (Kattge *et al.*, 2020) to characterize multiple scales of pattern without relying on covariate interpolation. PRISMA scenes are close to a random sample of tropical forests and are individually large enough to each encompass significant local topographic, soil and disturbance patterns. Remote and in situ trait measurements differ from each other in profound ways, sampling leaves versus pixels, plots versus scenes, and consistently the upper canopy versus foliar sampling reflecting restrictions to canopy access. As such, comparisons reveal a great deal through their differences, rather than one falsifying the other or vice versa (Wright *et al.*, 2004).

## Materials and Methods

### The PRISMA instrument

In 2019, the Italian Space Agency launched the PRISMA mission (Buongiorno *et al.*, 2021) to obtain hyperspectral imagery globally from space, acquiring 30 x 30 km scenes with 30 m spatial resolution, capturing targeted individual scenes, not wall-to-wall coverage.

PRISMA's 239 channels span the range from visible through shortwave-infrared, from ~400 through 2500 nm.  PRISMA's characteristics are shown in Table 1.

Table 1. Characteristics of the Italian Space Agency's PRISMA instrument.

| PARAMETER | VNIR CHANNELS | SWIR CHANNELS |
|---|---|---|
| Spectral Range (Nm) | 400-1010 | 920-2505 |
| Spectral Resolution | 9-13 nm | 9-14.5 nm |
| Number Of Channels | 63 | 171 |
| Signal-To-Noise Ratio (SNR) | >160 | >100 |
| Ground Sampling Distance (Pixel Size) | 30 m | 30 m |
| Swath Width | 30 km (1,000,000 pixels/scene) | |

## Scene and pixel selection

PRISMA scenes for tropical forests were downloaded from the Italian Space Agency and were selected to lie within the tropical rainforest, selected using a mask (Buchhorn *et al.*, 2020). Images were collected between 2019 and 2022. We sampled year-round (in tropical regions). Scenes were screened for clouds, and the vegetation fractional cover was calculated after spectral unmixing (Ochoa *et al.*, 2025) . We retained pixels with high vegetation cover (>90%), inferred after spectral unmixing of generic bare soil, green vegetation, dead vegetation, rock and water end members compiled from the EMIT database (https://ecosis.org/). The remotely sensed data capture known macro-climatic, hydrological and orographic gradients, as well as more local soil and disturbance gradients.

Cloud-free pixels were analyzed combining wavelengths from PRISMA's VNIR and SWIR sensors and associating each pixel to with its elevation (https://doi.org/10.5270/ESA-c5d3d65). We corrected spatial offsets by co-aligning PRISMA scenes with coincident Landsat 8 NIR imagery (Scheffler *et al.*, 2017).  Due to climatological cloud cover patterns, most scenes were obtained during the dry seasons.

Figure 1 shows the PRISMA scenes in relation to biomes and topography, while Supplemental Figure S3 shows the scenes in relation to geographic (latitude and longitude) and climate (mean annual temperature and rainfall).  The data provide dense sampling of the geographic and climate space.

While PRISMA's achieved signal-to-noise ratio (SNR) often exceeds the nominal values shown in Table 1, confident retrieval of plant traits requires high SNR and so we created 4 x

4 pixel "superpixels" by co-adding and averaging pixels together (Raiho *et al.*, 2023a).  This increases SNR from the values shown in Table 1 to closer to ~600 in the VNIR and ~400 in the SWIR, values known to enable plant functional trait retrievals (Raiho *et al.*, 2023a). While PRISMA's achieved SNR often exceeds the nominal values shown in Table 1 (Buongiorno *et al.*, 2021), confident retrieval of plant traits requires high SNR (Raiho *et al.*, 2023a).  The larger superpixels are also suited for the regional scale of the analysis, as key information about landscape variation and turnover is preserved for larger scale analyses (Ordway *et al.*, 2022).

## Trait Retrieval Methods

Retrieving plant functional traits starting with spectral radiances requires a number of preliminary image processing steps. First, radiometric correction was implemented via forward modelling (Guanter *et al.*, 2009) of a reference spectrum (assumed to be a homogeneous, temporally invariant and spectrally smooth site) to the top-of-atmosphere. Wavelength calibration coefficients were calculated for three PRISMA acquisitions over the reference site, representing the across-track wavelength shifts, minimizing the root-mean square error (RMSE) between observed and modelled reflectance for each wavelength. After the wavelength calibration, we derived radiometric corrections for each image as a ratio between observed and modelled radiances. The corrections were averaged from the three PRISMA acquisitions and applied to all scenes in this work. We used atmospheric corrections algorithms and radiative transfer models (Guanter *et al.*, 2009; Thompson *et al.*, 2021) to derive surface reflectance.

We calibrated an empirical model to retrieve plant canopy traits on the most comprehensive data available, aiming at a general model robust across diverse ecosystems, spanning a wide range of plant chemical, structural and canopy architecture strategies. Very limited data to directly parameterize a model for tropical forests is available, and instead we chose a calibration data set than spanned a wide range of plant chemical, structural and architectural strategies. The general model captures differences among grasslands, deciduous and evergreen forest, represented by both flowering plants and conifers, potentially capturing globally observed trait ranges (Díaz *et al.*, 2016; Serbin *et al.*, 2019; Wang *et al.*, 2020). We note that tropical forests include diverse chemical and structural features, not all of which are represented in the one available tropical training data set in Hawai'i (Table 2).

We retrieved the Chlorophyll, and $N_{mass}$ and LMA for PRISMA scenes using partial least square regression (PLSR) models trained using temperate to Hawai'ian tropical species in the National Ecological Observatory (NEON) network (Keller *et al.*, 2008). We also indexed canopy photosynthetic nitrogen use efficiency (pNUE) from the ratio of chlorophyll to nitrogen mass (Poorter & Evans, 1998; Pons & Westbeek, 2004).

We selected NEON sites with spatially coincident PRISMA scenes and seasonal matches (not always in the same year, but matched seasonally). We chose not to base the calibration entirely on Hawai'i as Hawai'ian forest diversity is low compared to the pan-tropics and does not span the full range of foliar structural and chemical characteristics (Westerband *et al.*, 2021).

Table 2. The NEON sites, ecosystem types and locations. These sites were used for calibration and testing of trait retrievals span a wide range of ecosystem types and physiognomies (Keller *et al.*, 2008).

| SITE | ECOSYSTEM TYPE | LATITUDE AND LONGITUDE |
|---|---|---|
| Konza Praire (Kansas) | Tallgrass Prairie | $39.1°, -96.6°$ |
| Niwot Ridge (Colorado) | Subalpine Forest | $40.1°, -105.6°$ |
| Oak Ridge National Laboratory (Tennessee) | Eastern Deciduous Forest | $35.9°, -84.3°$ |
| Ordway-Swisher Biological Station (Florida) | Coastal wetlands | $29.6°, -82.0°$ |
| Pu'u maka'ala Natural Areas (Hawai'i) | Hawai'ian Tropical Forest | $19.5°, 155.3°$ |
| Soaproot Saddle (Sierra National Forest, California) | Mixed Conifer Forest | $37.0°, -119.2°$ |
| Talladega National Forest (Alabama) | South Appalachian Mixed Deciduous Forest | $33.0°, -87.4°$ |
| University Of Notre Dame Environmental Research Station (Wisconsin) | Northern Mesic Hardwood Fiorest | $46.2°, -89.5°$ |
| Wind River Experimental Forest (Washington) | Old-growth Douglas Fir-Western Hemlock Coniferous Forest | $45.8°, -122.0°$ |

We generated trait maps for NEON sites after methods and datasets in Wang et al. (2020); this approach applies permutations of partial-least squares regression based on coincident canopy reflectance spectra obtained from the NEON Airborne Observatory Platform (NEON-AOP; 1m resolution) and site-level field trait data. The NEON trait maps were averaged to match the 120m resolution and alignment of the PRISMA reflectances and then intersected with the PRISMA imagery to generate a training dataset. We retained

PRISMA pixels with >90% vegetation. The training data were split with 50% for calibration and 50% reserved for validation, and we generated 100 permutations of PLSR models to predict NEON-derived leaf mass per area and nitrogen as a function of PRISMA spectra, drawing at a random 70/30 split between calibration and validation from the training datasets. The estimated 1-σ precisions of LMA, N and chlorophyll have been estimated by NASA's SBG mission researchers as 25%, 24% and 20% (Raiho *et al.*, 2023b) (Thompson, pers comm).

## Evaluating bias and uncertainty in PRISMA-retrieved traits

Trait retrievals were validated using the reserved pixels from the training data set (see above). Supplemental Figures 1 and 2 show the trait distribution of the training dataset for leaf mass and Nitrogen per unit dry mass and illustrates the comprehensive nature of this dataset, which includes grasslands and coniferous forests as endmembers. Supplemental Figures 1 and 2 show an x-y plot of predicted vs. observed trait estimates for pixels used for calibration and not included in the calibration. Calibration and validation results were similar for chlorophyll.

No contemporaneous for spectroscopic trait retrievals in tropical forests exist for direct ground truth. Two comparable, though not spatially or temporally coincident, two data sets are available in Peru (Asner *et al.*, 2017a; Kattge *et al.*, 2020). We used these data to ensure that the range and central tendency of retrieved traits was consistent with available leaf-level measurements for the tropics. We compared the central tendency and distribution of traits observed in Peruvian rainforest for PRISMA retrievals to two extant *in situ* compilations. We selected pixel-level traits (see previous details) from seventeen PRISMA scenes occurring in the Peruvian lowlands for comparison with co-occurring sources. For each scene, we randomly sampled twenty-nine thousand pixels (without replacement), representing the minimum pixel number recorded among all scenes. We excluded occurrences of anomalously low chlorophyll content and excluded all outliers above a z-score of 3 as these result from incompletely corrected atmospheric effects (Townsend, pers comm).

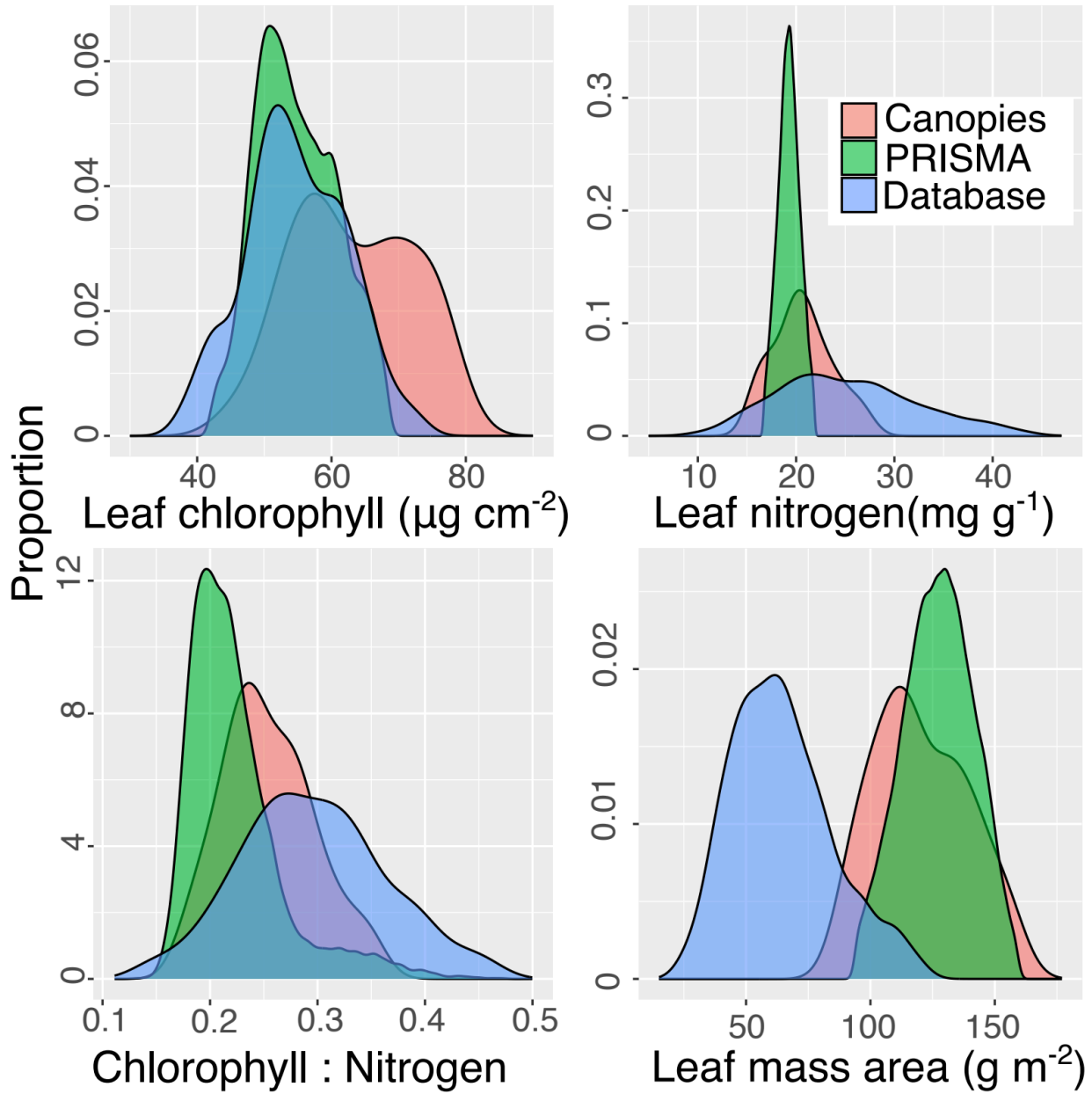

Figure 2. *Comparisons in Peru of PRISMA retrievals versus two in situ data distributions. The data are not direct ground truth but rather represent non-coincident samples from the population of traits within Peruvian moist forests.* (Asner *et al.*, 2017b; Kattge *et al.*, 2020). *The most comparable data (Canopies), sampling the upper canopy, shows good agreement given the data are not coincident is space or time, TRY data generally show broader distributions, reflecting their much more heterogenous nature.*

Peru holds one of the largest collections of plant traits from several years of intensive measurements. To compare PRISMA retrievals to the TRY database (Kattge *et al.*, 2020) of plant traits, we identified species found in Peru and then searched a the TRY data base for occurrences of those species in Peru. We built a species checklist for Peru querying the Kew plants of the world online resource (Govaerts *et al.*, 2021; Brown *et al.*, 2023), selecting arborescent species in moist tropical and seasonally dry tropical climates (resulting in 4143 unique species). We then searched for these species in a TRY table of taxonomically homogenized global trait records. The TRY trait query included all public records for specific leaf area, Nitrogen and Chlorophyll content and concentration (for 292, 020 accessions). This number of records was reduced to 32,456 records in 1300 species occurring in Peru; trait summaries were computed based on all records.

We also compare to a more directly comparable data set of in situ measurements on top-of-the canopy-sampled trait averages for 80 0.1 ha forest quadrants, collected in the Peruvian eastern lowlands, with top of canopy traits averaged by species abundance (Asner *et al.*, 2017b). We note that while Asner (Asner *et al.*, 2017b) trait estimates are always for the upper sunlit canopy, records in TRY are not necessarily from samples collected in the overstory.

The overall comparison establishes the credibility of the PRISMA canopy retrievals as corresponding to in situ-measured foliar traits (Fig. 2). The PRISMA retrievals generally agree with the Asner (Asner *et al.*, 2017b) ranges, even though they were not spatially co-located and were collected over a decade apart. The TRY ranges also roughly correspond, though not as closely, as might be expected given their greater heterogeneity in sampling, canopy position and methodology. Unlike LMA and chlorophyll content, nitrogen concentrations derived from PRISMA show lower variance when compared to the two other sources (Fig. 2b).

Were the PRISMA retrievals misleading or grossly biased, we would not expect close correspondence, and the largest discrepancy, with LMA, may be due to the mix of canopy positions sampled in TRY. Based on the Peru comparisons, and the extensive NEON calibration and validation, the PRISMA data appear to be a plausible representation of tropical forest traits, and we proceed to analyze them.

## Leaf economic spectrum and photosynthetic nitrogen use efficiency index

For computational feasibility, estimates of trait correlations LES and pNUE) between continents were based on a dataset including only five thousand pixels per scene (a 12.5% sample of 120 m pixels), this number of pixels is chosen to allow an equal representation of data for all scenes; for a few scenes that had less than five thousand pixels after screening for fractional cover and vegetation type, the resampling approach was done with replacement; otherwise, pixels were randomly chosen without replacement. With this dataset, we performed ordinary least squares regression to detect trait covariation and principal component analysis.

## Functional diversity

We estimated functional diversity for each scene and for each of the three continents, adapting the functional richness (FR) metric following (Schneider *et al.*, 2023). We base our estimate on standard deviation-scaled LMA and nitrogen concentration and calculate functional richness as as the area of a convex hull encompassing (all) the minimal area(s) capturing 75% of the datapoints and based on a kernel density map using the eks R package (Duong, 2024). This approach helped to limit our analysis to a subset of the distribution, avoiding noise in trait retrievals influencing functional richness estimates. As the number of vegetation pixels can vary substantially between scenes, and the number of scenes can vary between continents, we approach FD using a bootstrap method, selecting scenes for each continent at incremental numbers from 2 to 100.

For each selected scene, we selected randomly five thousand pixels, repeating it for 500 combinations, computing functional diversity. We then estimated the relationship between the number of combined scenes and the functional diversity, analogous to a species-area curve. We plotted functional diversity against area (# of scenes aggregated x functional diversity for that number of scenes) to visualize how functional diversity accumulates with additional sampling in the three continents. We chose to use number of scenes rather than distance between scenes as distance is ill-defined given the huge spatial areas and the fragmented nature of Asian land masses and between South and Central America, where much of the inter-plot distance would be ocean.

To relate topography to functional diversity, we computed the standard deviation of elevations associated with the center points of each 120 m pixel within scenes and used that as an index of topographic variation. We recognize that topographic variation only indexes some of the effects mountains can have on functional diversity (Kreft & Jetz, 2007), as orographic isolation of populations, rain shadows and other influences can play a role, and can be investigated in the future.

## Results and Discussion

The three tropical continents show somewhat different, though strongly overlapping LES trait distributions (Fig. 3). The plots show continental variation averaging over local or landscape scale. Examining central tendencies and ranges, the three continents do not occupy the same trait space, with overall amplitude always highest in the Americas, and always lowest in Africa. The continents differ rather strongly in distributions for LMA and chlorophyll, and are more similar for nitrogen content, though the Americas exhibit a dramatically wider range than the other continents. The differences between the continents become even more apparent in the LES and pNUE relationships (Fig. 4). The impacts of the differing trends in foliar N and chlorophyll are also evident in the plots of the LES axis, and the LMA-pNUE relationship (Fig. 4), suggesting different strategies may dominate in different parts of the biome.

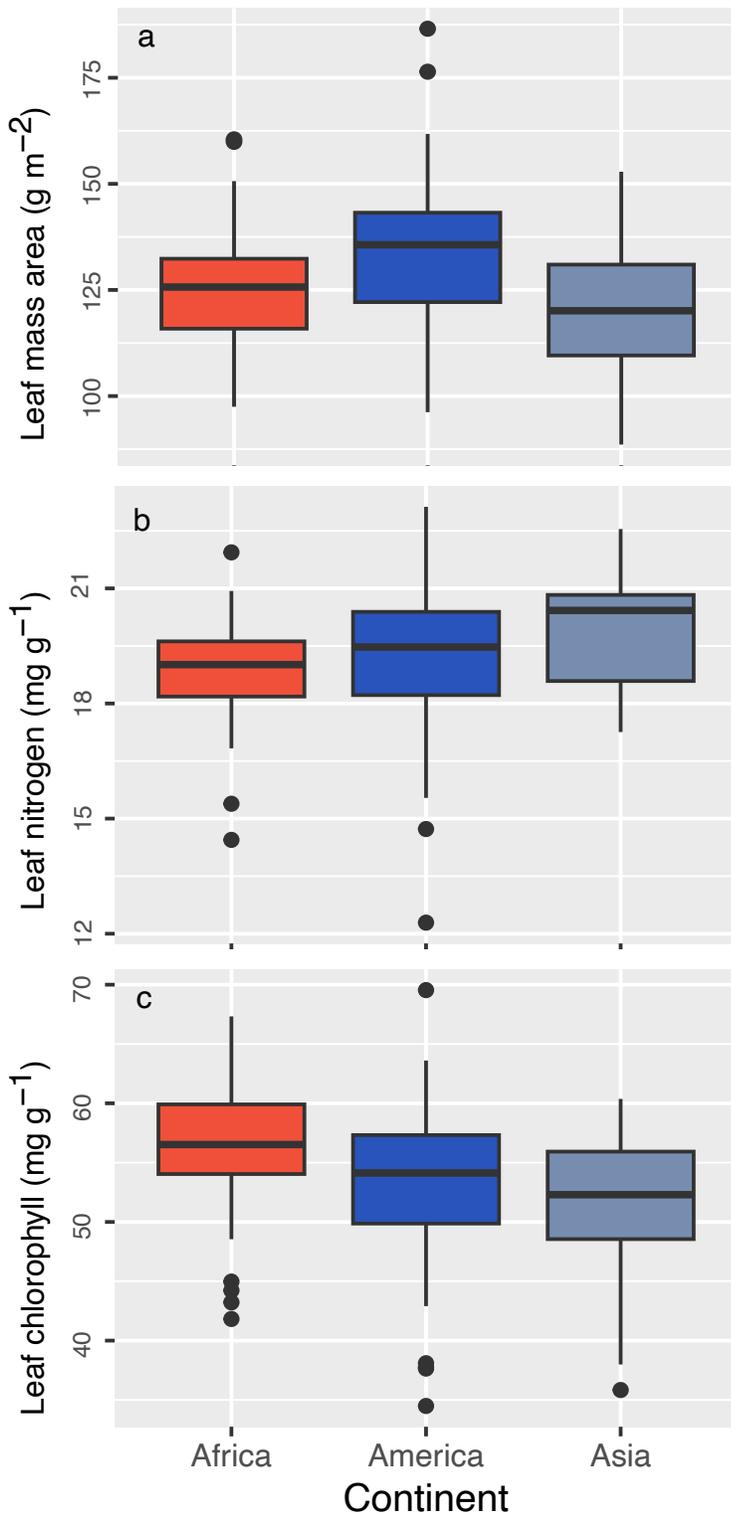

*Figure 3. Continental univariate distributions of traits differ by continent, top to bottom: Leaf mass per unit area, Leaf nitrogen and Leaf chlorophyll.  a-c LMA, leaf nitrogen concentration, and leaf chlorophyll content.*

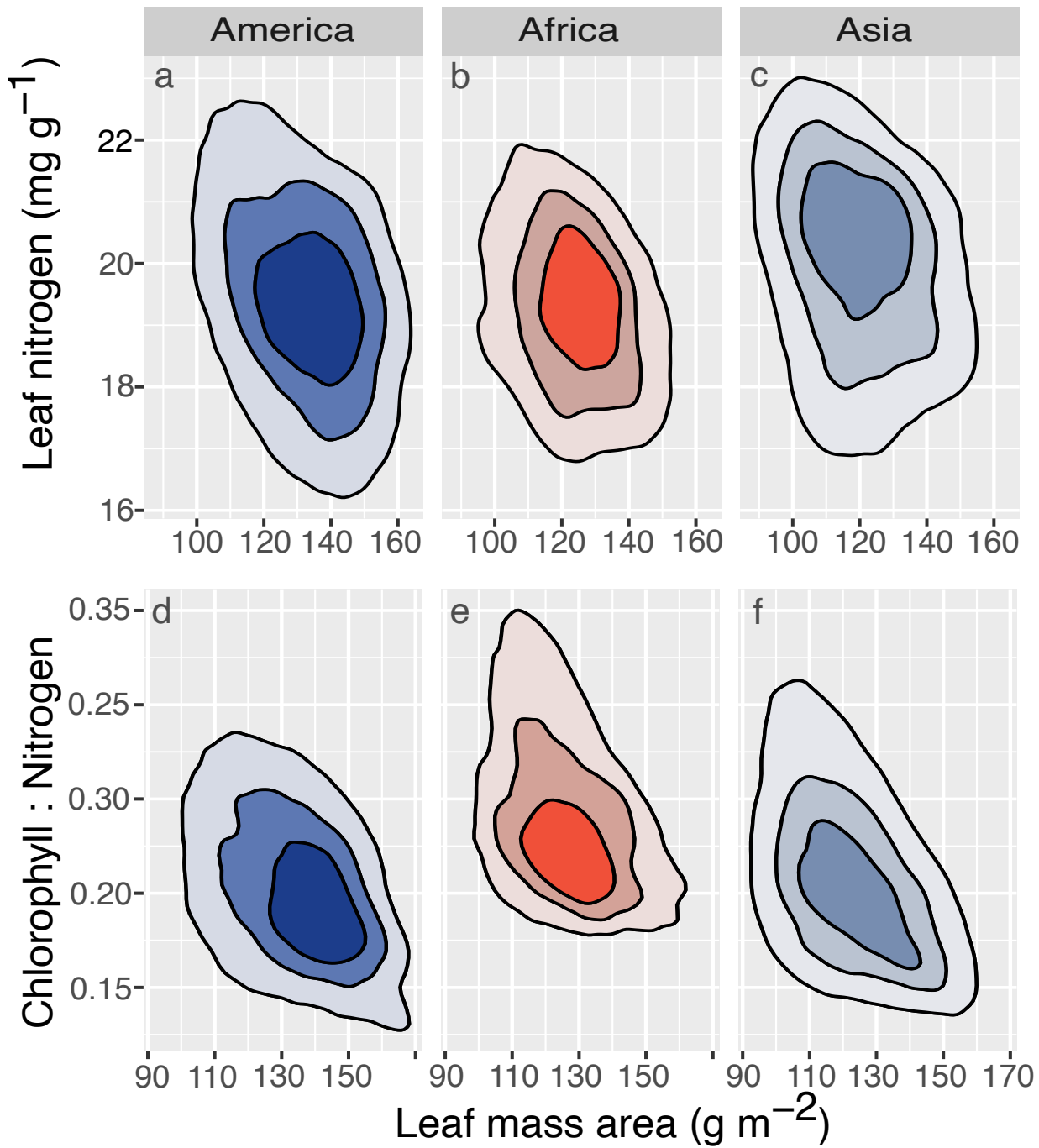

*Figure 4) The space defined by the LES varies between continents. Bivariate density plot of the LES (Leaf nitrogen versus leaf mass per unit area) for a-c, the Americas Africa, and Asia. The Americas and Asia have more amplitude along the LES axis than Africa. B) d-f Americas, Africa, Asia. The chlorophyll to nitrogen ratio (Photosynthetic nitrogen use efficiency, pNUE) indexes investment of a key nutrient in growth. All regressions are highly*

*significant (<0001) as would be expected from the large sample sizes, but the $R^2$ values (5% to 30%) are lower than seen in cross-biome analyses spanning wider (e.g., (Wright et al., 2004))(see supplemental Table 1).*

The three continents have somewhat different LES relationships (Fig. 4 a-c). LES trait correlations show the slope and amplitude of the LMA:Nitrogen and LMA:pNUE relationships. All three continents show clear LES and pNUE relationships. The slopes are similar, but the amplitudes differ, with Africa showing a significantly lower amplitude for N, while Asia shows the highest N values. Asia and the Americas have similar ranges for N, shifted with the Americas how lower maximum and minimum N values.

pNUE is an indicator of investment in growth (Onoda et al., 2017), as opposed to other foliar uses of nitrogen, such as defense compounds or reproduction. The three continents again differ, with Africa showing somewhat higher investment of N in chlorophyll, with Africa and the Americas being lower. Africa, lacking the very high N canopies seen for Asia and the Americas in Figure 4 a-c may compensate ecologically by allocating proportionately more N to growth (high pNUE: Figure 4 d-f).

Low pNUE (more investment in N in non-growth related compounds) occur at a wide range of LMA values, suggesting a range of strategies in play, possibly reflecting differing defense strategies (mobile versus static) (Coley et al., 1985). While the Americas show a wide range of LMA values at intermediate pNUE, Africa shows a wide range of LMA values at low LMA. The results confirm the generality of the LES relationship, while showing nuanced differences between the three continents in allocation of nitrogen to photosynthetic machinery as opposed to other uses.

The trait distributions between the continents are quite different from analyses "upscaling" from sparse in situ samples (Aguirre-Gutiérrez et al., 2025). The area sampled by the satellite, to the extent that one can compare plot and remote observations, represents approximately 20,000 times more of the overall moist forest biome. The data also differ in that reflectance-based measures are exclusively from the upper sunlit canopy while many situ data compilations include over-and-understory measurements (Aguirre-Gutiérrez et al., 2025). The impact of this can be visualized in Figure 2 (see Peru comparison) where the Asner plot data, collected from the upper canopy, agree in distribution much more closely than the broad TRY compilation, which includes many samples of opportunity, often not identified by where in the canopy they were collected.

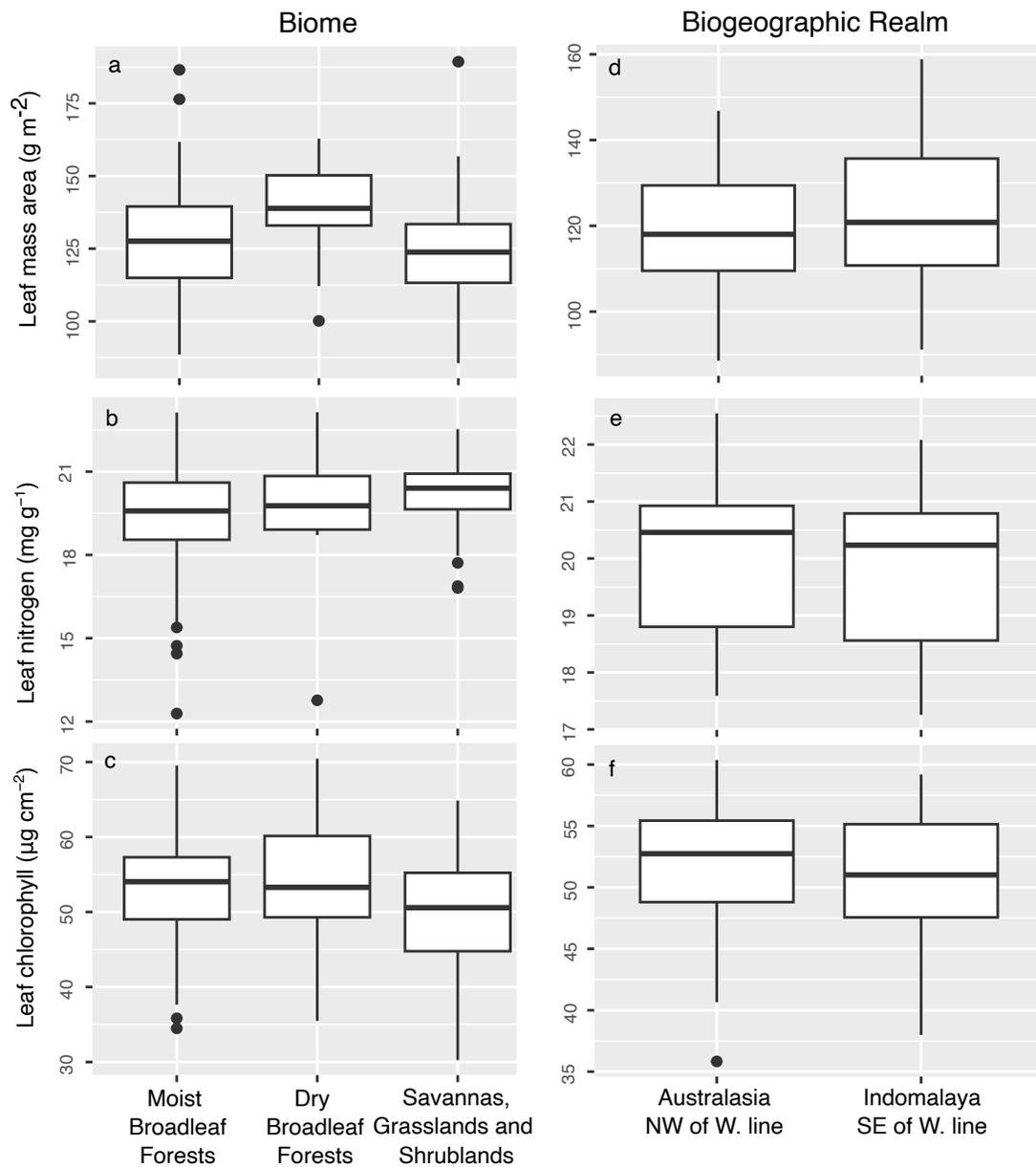

*Figure 5) a-c: Functional trait in moist tropical forest compared to adjacent tropical biomes. The three biomes lie in different parts of trait space, in addition to having differing physiognomy. d-f: Within the Asian moist forest, comparing the Australsian to the Indo-Malayan tropics east and west of Wallace's line. Both comparisons show significant variation.*

This paper focuses on tropical moist forests, but the overall PRISMA data set spans multiple tropical biomes. Moist forests are clearly structurally and climatically very different from the other, drier biomes, but they are also clearly offset in trait space. The three biomes we examined, moist forest, dry forest and grass/shrublands differ in leaf mass, nitrogen and chlorophyll, and also differ in their LES relationships. Patterns are not consistent between biomes, considering leaf mass, nitrogen and chlorophyll together (Fig. 5). Wet and dry forests are offset significantly from each other on the LES, with a trend towards heavy, long-lived leaves in dry forests. Grasslands are also offset, and this pattern may be influenced by the traits linked to their distinctive plant lineage photosynthetic habits. Tropical grasslands and shrublands have large C4 photosynthetic components with different resource use strategies (Ehleringer & Cerling, 2002), with light leaves compared to tree-dominated biomes. Surrounding grasslands could potentially influence the trait retrievals when forests are restricted to relatively patchy environments in a large grassland matrix.

Asia, as a series of islands and disjunct land masses and is often divided along Wallace's line, demarcated ecologically by very different fauna, while being similar in flora (Corlett & Primack, 2011). We examined trait distributions to either side of Wallace's line, as this is the first trait data set with sufficient sampling to zoom in to that degree (Fig. 5 d-f). We note rather significant differences between the Indo-Malaysian and Australasian trait distributions, despite taxonomic similarity. In Australasia, we find heavier leaves, lower nitrogen and slightly lower chlorophyll, suggesting an overall LES bias towards slower-growing strategies. Although beyond the data, the differences in foliar characteristics also influence palatability and food quality, and may also respond to long-term evolutionary pressure from herbivores.

Within the continents, topography may play a major role in trait variation. Figure 6 shows plots of scene-level variation and the range of elevation within a scene. Previous work in the tropics shows that orography, with attendant edaphic, hydrological and disturbance regimes, can lead to environmental filtering and that canopy traits can reflect resource availability along hillslopes (Chadwick & Asner, 2020). As a result, high $\beta$ diversity (trait turnover) may be expected along topo-edaphic resource gradients. These plots (Fig. 6-7) show a clear relationship between the range of variation in elevation within scenes and those scenes functional diversity, similar to (Kreft & Jetz, 2007). Asia is the most mountainous of the tropical regions and has scenes with correspondingly higher functional diversity. While previous global and pan-tropical studies have focused on continental scales of variation in traits, remote sensing allows consideration of both scene and larger scale patterns because of the density of observations (Dahlin *et al.*, 2013).

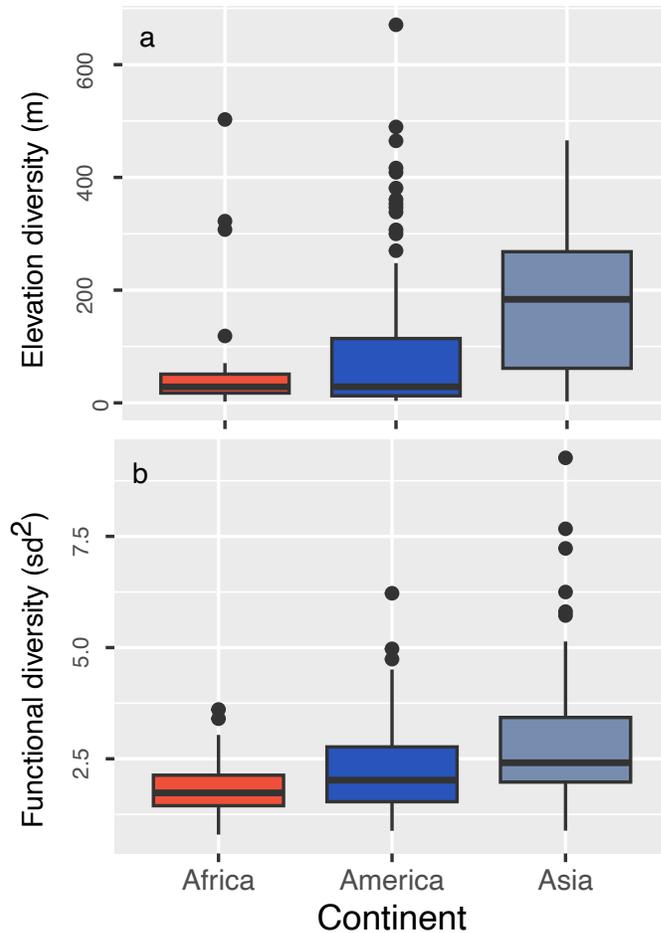

*Figure 6. Orography is an important correlate of FD. Elevation (a) and functional diversity (b) by scene, estimated from the convex hull around the LES traits. FD differs between Africa, the Americas and Asia. The mean and range of FD vary with to the mean and range for elevational diversity (the standard deviation of elevation within each scene).*

Mountains are a control on functional diversity. Across the scenes considered, orography (the standard deviation of elevation) seems to explain about half the variation in functional diversity (Fig. 7), suggesting climate, disturbance, herbivory and other soil-related parameters all also explain about half of the variation. Topography and trait-topography relationships should be factored into modeling and extrapolating trait distributions, and sampling must be adequate to capture these relationships in such models. Very high variation around the regression at low elevational variation may occur along rivers, where steep environmental gradients can occur from river margins to interfluves (Maracahipes-Santos *et al.*, 2023).

The spectrum of local-scale or landscape scale variation within the tropics is high, and results in challenges to scaling or extrapolation from limited and in situ estimates. Even careful random sampling within a sampling plot, or group of plots could alias local variability into estimates of large-scale variation if the number of replicates is insufficient to span the universe of variation in that landscape.

This may be mitigated by including covariates, for example canopy reflectance or vegetation height. Indices from low-dimensional sensors (Raiho *et al.*, 2023a) such as

Landsat and Sentinel may not capture landscape variation if it is not reflected in the indices employed. Most multispectral indices are linked to light harvesting, photosynthesis, and overall vegetative cover (Aguirre-Gutiérrez *et al.*, 2025), which may or may not be correlated with other LES traits, much less those linked to reproduction or defense (Serbin & Townsend, 2020).

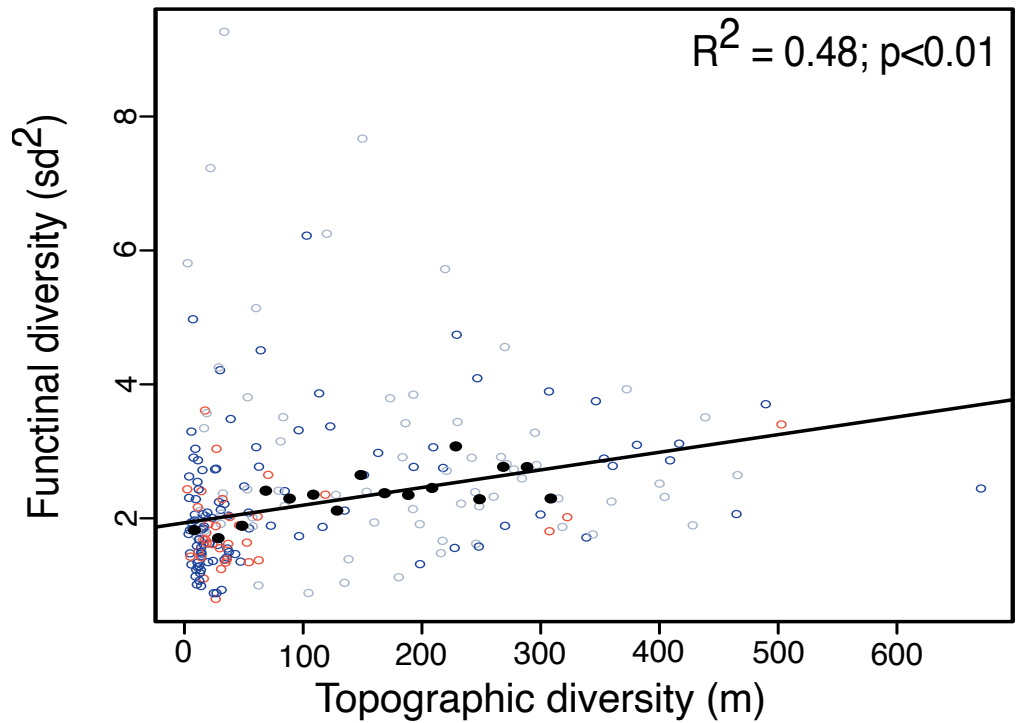

*Figure 7. A linear relationship explains about half of the variance in the relationship between FD and elevational diversity. Trait correlations with macroclimate and other factors (soils, disturbance) are expected so the regression is on smoothed medians of FD, and considerable variance around this relationship occurs especially at very low FD, likely representing gradients in scenes that span river margins, flooded areas and uplands with low relief. Colors are red for Africa, Blue for the Americas and gray for Asia. Solid points are medians for 100 m elevation bands.*

Different processes may dominate between scales of trait variation. Landscape filtering and adaptations increase the spectrum of variation available to enable species replacement and migration. However, specialization along topo-edaphic or disturbance gradients (Ordway *et al.*, 2022) may also increase vulnerability if species are highly specialized and their local neighborhood conditions, or their entire fundamental niche, are not found in the new environments. While knowing the spectrum of variation is critical for predicting ecosystem response in a changing environment (or for explaining changes in the past), the "devil is in the details" and overall variation should not be automatically associated with resilience in the face of stress. Africa provides an example of this, with substantially lower overall LES variation, yet stability of biomass over the past decades, in contrast to more diverse regions.

Functional diversity (Fig. 8) shows similar scaling to estimates of forest species diversity (Slik *et al.*, 2015). To compare to (Slik *et al.*, 2015) we calculated the equivalent of species-area curves, accumulating functional diversity as more and more scenes are combined. The results show continental relationships that parallel those shown for species diversity, with the Americas and Asia having higher functional diversity, as (Slik *et al.*, 2015) showed for species diversity. Functional diversity saturates at a similar number of combined scenes in all three continents. Africa has significantly lower species diversity (Slik *et al.*, 2015), and in parallel, lower functional diversity (Fig. 8). Africa has both less univariate trait variation and lower amplitude along the LES-related axes (N and pNUE versus LMA), captured in its overall lower functional diversity.

At the scale of single scenes, Asia is (on average) more functionally diverse than America (Fig. 6b), but at a continental level these differences are not compounded (Fig. 8). The area of a biome should influence overall diversity (Hubbell, 2011; Raven *et al.*, 2020), but here we find a contrasting pattern; the much larger moist forest biome in America converges to similar estimates of functional diversity as the overall smaller Asian region (Fig. 8). The reasons why these patterns should be convergent is unknown, inviting future exploration of the evolutionary and environmental adaptations that underly constraints on functional diversity.

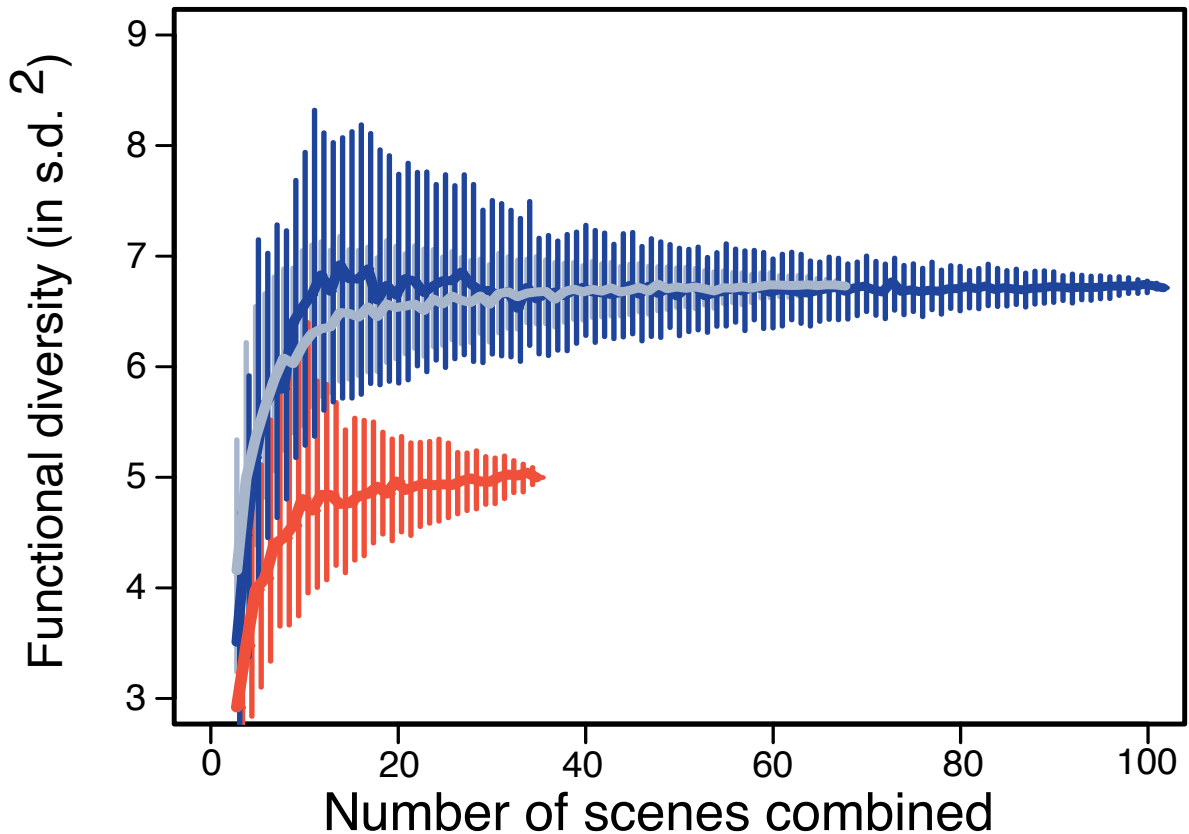

*Figure 8. Functional Diversity-area curves by continent show scaling, saturating around 10-20 scenes combined (9,000-18,000 sq km) but different levels of FD. African and American tropical forests have similar and high FD, while Africa is much lower. FD saturates while in contrast reported measures of species diversity continue to increase* (Slik *et al.*, 2015). *Red = Africa, Blue = The Americas, Gray = Asia*.

## Conclusions

Africa, Asia and the Americas each occupy a different part of the space defined by the LES traits. Asia and the Americas have very wide trait and LES axis variation, and correspondingly high functional diversity, contrasting with patterns found in modeled distributions (Aguirre-Gutiérrez *et al.*, 2025). Africa varies less in individual traits and is even more constrained along the LES axis, while having the widest range on the pNUE versus LMA axis, potentially indicating species with very high growth rates and low investment in other strategies. Differences in the diversity of the LES traits parallel continental differences in tree species diversity (Slik *et al.*, 2015). The continents, while overlapping in growth related traits, are clearly not merely replicates of each other.

While trait distributions and variation undoubtedly reflect many influences, including climate, climate history, evolutionary origin and the many ecological differences, differences in landforms appear to contribute significantly.  Mountains create a range of environmental conditions, as well as creating isolation between populations, and so can lead to increased species and functional diversity (Donoghue, 2008).  Remotely sensed plant functional traits are associated with orography, with a strong relationship between topographic variation and functional diversity.  This study analyzes approximately 5 orders of magnitude more area than extant plot-based studies (Aguirre-Gutiérrez *et al.*, 2025) and so has far more power to capture the multiple scales of trait variation  across the pan-tropics (Messier *et al.*, 2010).

Overall, topographic variation explains about half the observed global range in functional diversity, suggesting that landforms play a major role in the development and maintenance of diverse strategies.  This aspect of functional variation will affect upscaled estimates, and appropriate analysis is needed to develop relationships between plant function and relevant covariates. The association of traits and their variation with mountain ranges will affect the range of strategies to be considered in modeling. The role of those strategies in ecosystem response to environmental change is unknown, with the diversity of traits potentially increasing the range of responses while specialization and dispersal barriers could lead to more stress (Chen *et al.*, 2025).

Amongst the many known differences between tropical biomes, continents and between the Australasian and Indo-Malaysian (Corlett & Primack, 2011) (E and W of Wallace's line) tropics are differences in higher trophic levels, and the types, numbers and sizes of herbivores (Corlett & Primack, 2011).  Tropical trees often invest in defense (Wang *et al.*, 2023), and so herbivory may play a significant role in  trait distributions, at scales defined by animal activity (Coley *et al.*, 1985).  The LES traits are linked to herbivory: N through protein content and food quality, and LMA through palatability and toughness., and so may interact with higher trophic levels (Cardenas *et al.*, 2014). Are some of the differences seen related to herbivory and differences in the recent and evolutionary role of herbivores?  This is an open question and may be addressable as technologies for space-based spectroscopy improve (Green, 2022).

Traits vary at multiple scales (Messier *et al.*, 2010).  While global and pan-tropical analyses suggest correlations with macro-climatic factors, soils and other factors, this analysis captures multiple scales of variation directly, from pixels within scenes, to scenes within continents spanning landform gradients and between continents.  There is considerable information from field studies about influences over trait distributions and this rich literature provides critical context for analyzing "big data" from compilations, and this should be considered in training models (Messier *et al.*, 2010; Chadwick & Asner, 2020).

Multiple scales of variation are important ecologically as well as for upscaling from limited local data. Upscaling will be impacted by sampling bias to certain landscape positions, often by accessibility, and by the availability of relevant covariates. If samples are not collected across landform gradients, the appropriate covariates may not even be evident, and so conclusions from spare sampling may address some, and not all relevant scales of variation. In any case, with orography explaining ~half of variation in functional diversity, macro-ecological variables should at best fill in the other half.

This study lays the groundwork for the global application of imaging spectroscopy to understanding ecosystem function through the lens of plant functional traits Cavender-Bares). These data provide information on potential plant growth performance, and future, higher performance sensors will provide information on a wider range of traits, including structural (eg lignin) and chemical (eg, glycosides) aspects of defense (Coley *et al.*, 1985). Sensors currently on orbit and planned will also allow monitoring change in plant function and functional diversity and will reveal how a changing environment interacts with plant strategies to influence trajectories of change.

## Acknowledgments

The research carried out at the Jet Propulsion Lab, California Institute of Technology, was under a contract with the National Aeronautics and Space Administration. Government sponsorship is acknowledged. This project was supported by a NASA-sponsored project # N4-CARBON24-0002 to JPL and the Goddard Space Flight Center. The research carried out at the University of Maryland by AB was under a contract (80NSSC23M0011) with NASA. We thank the Agenzia Spaziale Italiana (ASI) for support of this work, providing access to large amounts of data from the PRISMA instrument under its open use policy. Thanks to Jens Kattge (TRY), Greg Asner, Robin Martin and co-authors of the reported Peruvian canopy trait data used for comparison. Ryan Pavlick, while at JPL, made major contributions to the genesis of this research. Latha Baskaran, of JPl, provided the artistic touch for Fig. 1; we thank her.

# *New Phytologist* Supporting Information

Article title: Pan-tropical functional trait variation from space

Authors: David Schimel[1], Andres Baresch[2], Adam Chlus[1], Phil Townsend[3], Fabian Schneider[1,4], Gaia Vaglio Laurin[5] and Ben Poulter[2]

Article acceptance date: Click here to enter a date.

The following Supporting Information is available for this article:

**Fig. S1** *Diagnostics for PLSR models to predict LMA as a function of PRISMA spectra. Upper left: distribution of trait values used in model. Upper right: Selection of number of latent vectors in PLSR model based on minimization of the PRESS statistic across 100 data permutations. Middle left: performance of PLSR model on training data. Middle right: Performance of PLSR model on validation data. Different colors represent different NEON sites. Bottom: PLSR beta coefficients for trait predictions.*

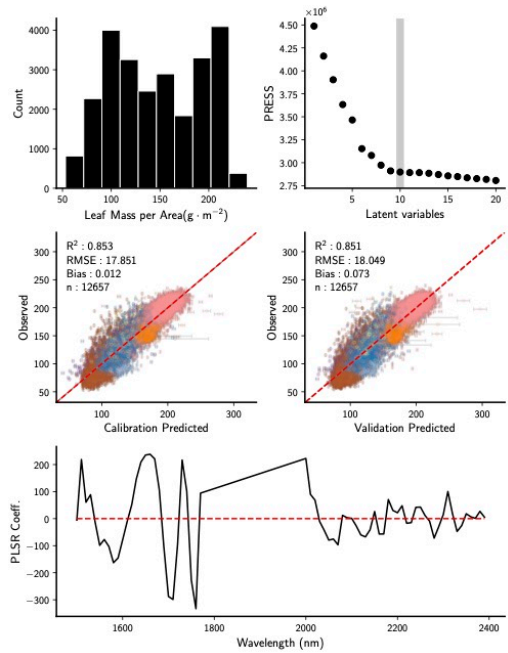

**Fig. S2** *Diagnostics for PLSR models to predict Nitrogen as a function of PRISMA spectra. Upper left: distribution of trait values used in model. Upper right: Selection of number of latent vectors in PLSR model based on minimization of the PRESS statistic across 100 data permutations. Middle left: performance of PLSR model on training data. Middle right: Performance of PLSR model on validation data. Different colors represent different NEON sites. Bottom: PLSR beta coefficients for trait predictions.*

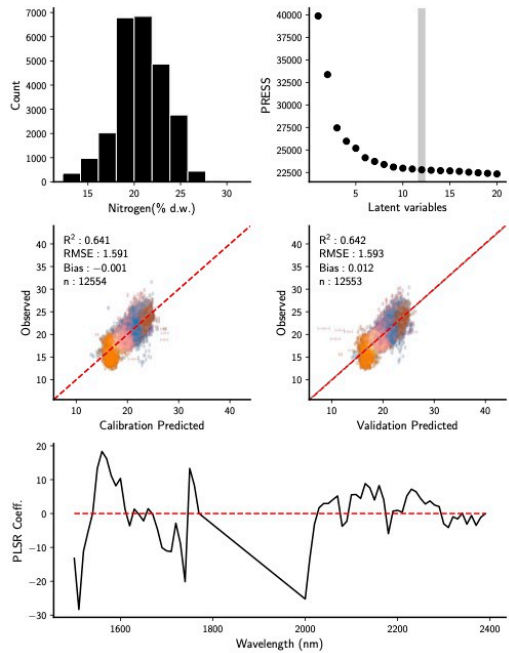

**Fig. S3** Figure S1. a) latitudinal and longitudinal distribution for the centroids in each of the PRISMA acquisitions considered in this work. b) climatic distribution of the PRISMA acquisitions. Average annual temperature (x axis) and annual precipitation (y axis) are derived from the CRU global climate dataset (at 0.5 º resolution) and sampled at the centroid of each PRISMA acquisition. Temperature and precipitation are an average for yearly values between 2010 and 2020.

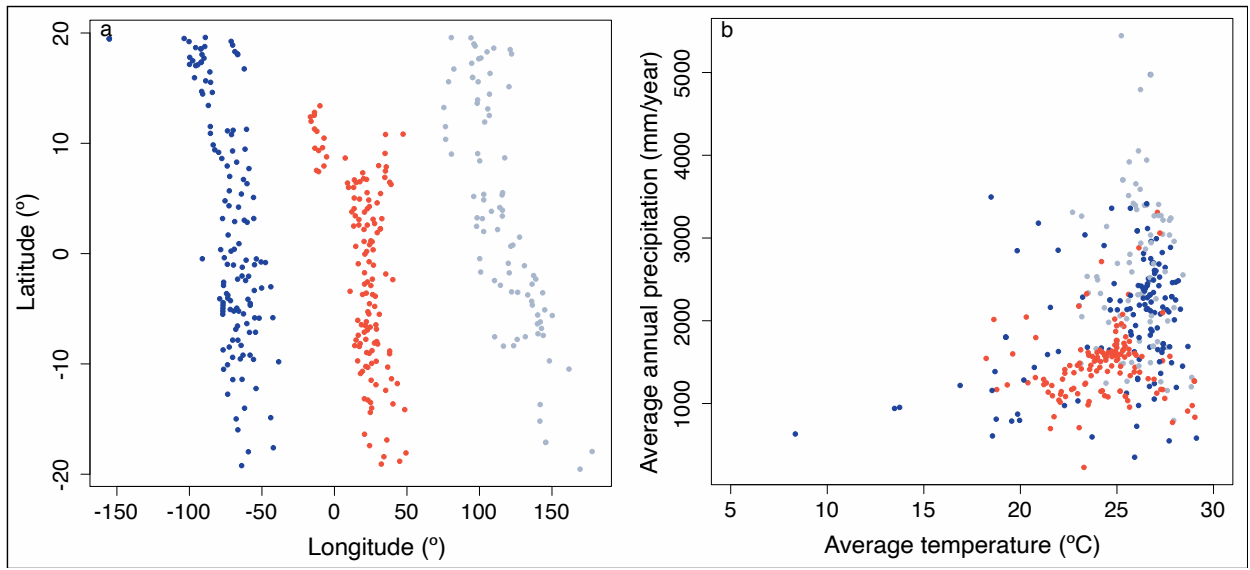

**Table S1. Regression statistics for Figure 4a-f computed over 120 m resolution pixels.**

LES regression (N vs LMA)

| Continent | model | Slope | y-intercept | R2 | Pval | n (pixels) |
|---|---|---|---|---|---|---|
| America | N as f of LMA | -0.0342117 | 23.917851 | 0.1247 | <0.0001 | 513905 |
| Africa | N as f of LMA | -0.0346204 | 23.5037049 | 0.1482 | <0.0001 | 167952 |
| Asia | N as f of LMA | -0.0201122 | 22.2659728 | 0.04718 | <0.0001 | 331652 |

pNUE regression

| | | | | | | |
|---|---|---|---|---|---|---|
| South America | chl:N as f of L | -0.0014381 | 0.4049101 | 0.2451 | <0.0001 | 432876 |
| Africa | chl:N as f of L | -0.001448 | 0.4271457 | 0.2245 | <0.0001 | 159955 |
| Asia | chl:N as f of L | -0.0018646 | 0.44878634 | 0.3145 | <0.0001 | 296587 |

**Data and software availability statement.** We follow open science guidelines:

1) Level 1 and 2 30m radiance and reflectance are available from ASI: https://www.asi.it/en/earth-science/prisma/
2) Level 3 (gridded reflectance) will be available at the time of publication at the Oak Ridge Distributed Active Access Center: https://daac.ornl.gov/
3) Level 4 (gridded plant functional traits) along with location and elevation, will be available at the time of publication at the Oak Ridge Distributed Active Access Center: https://daac.ornl.gov/
4) Software to produce reflectances are available at:

https://github.com/sister-jpl/sister

https://github.com/isofit/isofit